\documentclass[superscriptaddress,amsmath,amssymb,aps,prx,notitlepage,twocolumn,floatfix]{revtex4-2}
\usepackage[utf8]{inputenc}
\usepackage{listings}
\usepackage{footmisc}
\usepackage{enumerate}
\usepackage{latexsym}
\usepackage{braket}
\usepackage{graphicx}
\usepackage[caption=false]{subfig}
\usepackage[colorlinks=true, allcolors=blue]{hyperref}
\usepackage{blkarray}
\usepackage{array}
\usepackage[dvipsnames]{xcolor}
\usepackage[normalem]{ulem}
\usepackage{wasysym}
\usepackage{float} 
\usepackage{makecell}

\newcommand{\nk}[1]{\mathrm{#1}}

\begin{document}
\title{Improved effective vertices in the multi-orbital Two-Particle Self-Consistent method from Dynamical Mean-Field Theory}

\author{Karim Zantout}
\affiliation{Institute for Theoretical Physics, Goethe University Frankfurt, Max-von-Laue-Strasse 1, 60438 Frankfurt am Main, Germany}

\author{Steffen Backes}
\affiliation{Research Center for Advanced Science and Technology, University of Tokyo, Komaba, Tokyo 153-8904, Japan}
\affiliation{Center for Emergent Matter Science, RIKEN, Wako, Saitama 351-0198, Japan}
\affiliation{CPHT, CNRS, École polytechnique, Institut Polytechnique de Paris, 91120}

\author{Aleksandar Razpopov}
\affiliation{Institute for Theoretical Physics, Goethe University Frankfurt, Max-von-Laue-Strasse 1, 60438 Frankfurt am Main, Germany}

\author{Dominik Lessnich}
\affiliation{Institute for Theoretical Physics, Goethe University Frankfurt, Max-von-Laue-Strasse 1, 60438 Frankfurt am Main, Germany}

\author{Roser Valent\'\i}
\affiliation{Institute for Theoretical Physics, Goethe University Frankfurt, Max-von-Laue-Strasse 1, 60438 Frankfurt am Main, Germany}

\begin{abstract}
In this work we present a multi-orbital form of the Two-Particle Self-Consistent approach (TPSC),
where the effective local and static irreducible interaction vertices are determined
by means of the Dynamical Mean-Field Theory (DMFT).
This approach replaces the approximate ansatz equations for the double occupations $\langle n^{}_{\alpha,\sigma}n^{}_{\beta,\sigma'}\rangle$ by sampling them directly for the same model using DMFT. 
Compared to the usual Hartree-Fock like ansatz, this leads to more accurate local vertices
in the weakly correlated regime, and provides access to stronger correlated systems that were previously out of reach.
This approach is extended by replacing the local component of the TPSC self-energy
by the DMFT impurity self-energy, which results in an improved self-energy that incorporates
strong local correlations but retains a non-trivial momentum-dependence.
We find that this combination of TPSC and DMFT provides a significant improvement over the multi-orbital formulation of TPSC, as it allows to determine the components of the spin vertex without artificial symmetry assumptions, and opens the possibility to include the transversal particle-hole channel.
The new approach is also able to remove unphysical divergences in the charge vertices in TPSC. We find a general trend that lower temperatures can be accessed in the calculation.
Benchmarking single-particle quantities such as the local spectral function with other many-body methods
we find significant improvement in the more strongly correlated regime.
\end{abstract}

\maketitle

\section{Introduction}

The study of correlated electron physics in solid state systems poses many difficulties,
of which the solutions are expected to be key ingredients for understanding emergent phenomena such as unconventional superconductivity~\cite{Anderson1987,Kotliar1988,Lee2006,Honerkamp2008,Borisenko2010,WangScience2011,Norman2011,Jerome2012,Steglich2016,Aoki2019} or spin liquid phases~\cite{Balents2010,witczak2014,norman2016,Savary2017,Zhou2017,riedl2019,takagi2019}. 
As solving the many-electron problem exactly is impossible due to the large number of particles,
effective low-energy lattice models have been developed, such as the
Hubbard model, to describe the physics of correlated electrons in partially-filled orbitals~\cite{Hubbard1963,Kanamori1963,Gutzwiller1963}.
Even though this model is a significant simplification of the original problem,
in general an exact solution is not available, which led to the development of
a large variety of approximate many-body methods for the Hubbard model~\cite{Rohringer2018, Koch2019, Schaefer2021, Qin2022}. 
Among the different approaches we focus here on the Dynamical Mean-Field Theory (DMFT)~\cite{metzner1989,georges1992,Georges1996} and the Two-Particle Self-Consistent (TPSC) method~\cite{Vilk1997, Tremblay2012, Zantout2021}.

The idea of DMFT is to map the original lattice problem onto an effective local model,
embedded in an effective environment that is determined self-consistently.
This approach restricts correlation effects to purely local but dynamical contributions,
and thus results in a momentum-independent but frequency-dependent self-energy $\Sigma(\omega)$. 
This approximation becomes exact in the limit of infinite coordination number, but it has been applied to great success in finite-dimensional strongly correlated electron systems~\cite{Georges1996,Kotliar2004, biermann2005, Kotliar2006, georges2013, Vollhardt2019}. 
On the other hand, DMFT cannot describe non-local correlation effects, which are relevant e.g. in low-dimensional systems and the description of pseudo-gaps in the context of high-temperature cuprate superconductors~\cite{Norman1997, Ronning1998, Imada1998,Kordyuk2015}. 
Non-local extensions to DMFT that reintroduce a momentum-dependence to the self-energy have been and are actively developed,
such as cluster extensions~\cite{Georges1996,Hettler1998,Hettler2000,Maier2000,Kotliar2001,Maier2005,Park2008},
or different types of diagrammatic extensions such as 
GW+DMFT~\cite{Biermann2003,Sun2004,Ayral2012,Biermann2014,Boehnke2016,Backes2022},
dynamical vertex approximation~\cite{Toschi2007,Held2008,Galler2018,Rohringer2018},
TRILEX~\cite{Ayral2015,Ayral2016,Ayral2017}, 
QUADRILEX~\cite{AyralQUADRILEX2016}, 
D-TRILEX~\cite{Stepanov2019,Stepanov2021,Stepanov2021_2,Vandelli2022},
dual fermion~\cite{Rubtsov2009,Hafermann2009,Brener2020} and 
dual boson techniques~\cite{Rubtsov2012,Loon2014,Stepanov2016,Stepanov2016_2}.
Most of them come at a significant increase in computational cost, especially when multi-orbital systems are considered.

Another way to access non-local and dynamical correlation effects is given by the TPSC approach~\cite{VilkTremblay1997, Tremblay2012}.
Within TPSC the two-particle irreducible vertex $\Gamma$, which contains information on two-particle scattering processes, is approximated to be an effective local and static quantity~\cite{VilkTremblay1997}.
This effective interaction vertex $\Gamma$ is determined by requiring the resulting 
spin- and charge susceptibilities to fulfill corresponding local sum rules.
The quasi-particle renormalization effects from spin and charge fluctuations eventually lead to a non-local and dynamical self-energy $\Sigma(k,\omega)$.
In the case of weak to intermediate coupling strengths TPSC has proven to yield accurate results~\cite{VilkTremblay1997,Tremblay2012} and it can be extended to multi-site~\cite{Aizawa2015, Arya2015,Ogura2015,Zantout2018,Mertz2019, Pizarro2020} and multi-orbital~\cite{Miyahara2013,Zantout2019,Zantout2021} systems.
The multi-orbital TPSC formalism allowed for the investigation of possible spin-fluctuation pairing scenarios in unconventional superconductors~\cite{Miyahara2013} and non-local correlation effects in the spectral properties of Fe-based superconductors~\cite{Zantout2019,Bhattacharyya2020}, but still faces certain limitations, which we aim to address in this article.

One of the limitations of the multi-orbital TPSC scheme is the occurrence of diverging negative charge vertices when one requires the corresponding sum rules to be satisfied exactly, leading to unphysical negative spectral weight~\cite{Zantout2021}.
This issue can be circumvented by restricting the charge vertex search to non-negative values~\cite{Miyahara2013, Zantout2019, Zantout2021}, at the expense of violating the corresponding local sum rules.
Moreover, the usual ansatz equations do not allow for a determination of all double-occupations 
and correspondingly do not provide enough sum rules to determine all relevant elements of the spin vertex.
Symmetry relations that hold for the bare spin vertex can be used to determine these elements,
but still pose an additional approximation and do not allow for a straightforward inclusion of the transversal particle-hole channel~\cite{Zantout2021}.

In this paper we demonstrate that these limitations can be resolved by replacing the ansatz
equations for the double occupations with the double occupations directly obtained from a DMFT calculation.
This leads to effective local spin and charge vertices which are consistent
with the double occupations obtained by DMFT, and do not rely on the additional
Hartree-Fock like decoupling employed in the original ansatz equations.
Our results show that this approach, that we call $\langle nn \rangle^{}_{\text{DMFT}}$-TPSC, is able to improve or completely resolve the shortcomings of the original formulation of multi-orbital TPSC. 
The idea of using the double occupation as external parameter was already proposed in the original TPSC formulation~\cite{Vilk1997}.
In an additional step, the local self-energy of $\langle nn \rangle^{}_{\text{DMFT}}$-TPSC
can be replaced by the impurity self-energy of DMFT, which we name $\langle nn \rangle^{}_{\text{DMFT}}$-TPSC+$\Sigma^{}_\mathrm{DMFT}$.
This replacement is motivated by the non-perturbative nature of the DMFT approximation
which has shown to be able to obtain accurate results for local quantities such
as the local self-energy and double occupations~\cite{Georges1996, Kotliar2004,Schaefer2021}.
The same extensions of TPSC but in the single-band case are presented in Ref.~\cite{Martin2022}.
The article is structured as follows. In Section~\ref{sec: model_method} we present the multi-orbital TPSC formalism from Ref.~\cite{Zantout2021}, discuss the limitations and motivate the combination with DMFT. 
In order to test our new approach we perform calculations on a simple two-orbital Hubbard square lattice model as in Ref.~\cite{Zantout2021}. In section~\ref{sec: results}
the results of this benchmark are presented and discussed, and
Section~\ref{sec: conclusion} contains a summary and addresses open questions.

\section{Model and Method}
\label{sec: model_method}

In this work we consider the fermionic multi-orbital Hubbard-model, defined by the Hamiltonian
\begin{align}
H =& \sum_{\alpha,\beta, i, j,\sigma}\left(t_{\alpha\beta}^{\vec R_i-\vec R_j}-\mu\delta^{}_{i,j}\delta^{}_{\alpha,\beta}\right)c_{\alpha,\sigma}^\dag(\vec R_i)c^{}_{\beta,\sigma}(\vec R_j) \nonumber\\
+& \frac{1}{2}\sum_{\alpha,\beta,i,\sigma}U_{\alpha\beta}n_{\alpha,\sigma}(\vec R_i)n_{\beta,-\sigma}(\vec R_i)\nonumber\\
+& \frac{1}{2}\sum_{\substack{\alpha,\beta,i,\sigma \\ \alpha \neq \beta}}(U_{\alpha\beta}-J_{\alpha\beta})n_{\alpha,\sigma}(\vec R_i)n_{\beta,\sigma}(\vec R_i) \nonumber\\
-&\frac{1}{2}\sum_{\substack{\alpha,\beta,i,\sigma \\ \alpha \neq \beta}}J_{\alpha\beta}\left(c_{\alpha,\sigma}^\dag(\vec R_i)c^{}_{\alpha,-\sigma}(\vec R_i)c^{\dag}_{\beta,-\sigma}(\vec R_i)c^{}_{\beta,\sigma}(\vec R_i)\right.\nonumber\\
+&\left.c^{\dag}_{\alpha,\sigma}(\vec R_i)c^{}_{\beta,-\sigma}(\vec R_i)c^{\dag}_{\alpha,-\sigma}(\vec R_i)c^{}_{\beta,\sigma}(\vec R_i)\right),\label{Eq-Hubbard-H}
\end{align}
where $t_{\alpha\beta}^{\vec R_i-\vec R_j}$ are the hopping matrix elements between orbitals $\alpha$ and $\beta$ that are connected by lattice vectors $\vec R_i - \vec R_j$.
Here we restrict ourselves to the paramagnetic phase.
The Hubbard and Hund's coupling terms are denoted by $U_{\alpha\beta}$ and $J_{\alpha\beta}$ respectively.
Throughout this work we assume spherical symmetry with $U^{}_{\alpha\alpha}=U$,
and $U^{}_{\alpha\beta}=U-2J$ for $\alpha\neq \beta$.
The operator $c^{}_{\alpha,\sigma}(\vec R_i)$ destroys an electron with spin $\sigma$ in the $\alpha$-orbital at unit cell position $\vec R_i$ and $c^\dag_{\beta,\sigma}(\vec R_j)$ creates an electron with spin $\sigma$ in the $\beta$-orbital at unit cell position $\vec R_j$.
The density operator is defined via $n^{}_{\alpha,\sigma}(\vec R_i):=c^{\dag}_{\alpha,\sigma}(\vec R_i)c^{}_{\alpha,\sigma}(\vec R_i)$, and $\mu$ is the chemical potential.

In order to introduce the combined TPSC and DMFT approach to the multi-orbital Hubbard model, we first start with a short summary of the single-orbital and multi-orbital TPSC approach following Ref.~\cite{Zantout2021}.

\subsection{Single-orbital TPSC}

The Two-Particle Self-Consistent approach was originally developed for the single-band Hubbard model~\cite{VilkTremblay1997,Tremblay2012} based on an approximate expression for the Luttinger-Ward functional~\footnote{The Luttinger-Ward functional is a central functional in the Kadanoff-Baym formalism and defined as sum of all closed two-particle irreducible skeleton diagrams that can be constructed from the Green's function $G$ and the Hubbard interaction $U$.}, namely $\Phi[G]\approx G\Gamma G$, where $G$ is the full Green's function and $\Gamma$ is the two-particle irreducible vertex.
Additionally, in TPSC one approximates the irreducible vertex $\Gamma$ to be local and time-independent quantity.
The reasoning behind both approximations lies on the observation that far away from phase transitions higher order correlation functions only contribute through their averages which are assumed to be absorbed in an effective interaction vertex $\Gamma$.
In TPSC this effective interaction vertex $\Gamma$ appears as effective constant spin and charge vertices, $\Gamma^\text{sp}$ and $\Gamma^\text{ch}$ respectively.

In order to determine the values of the effective vertices $\Gamma^\text{sp}$ and $\Gamma^\text{ch}$, TPSC relies on the enforcement of the local spin and local charge sum rules
\begin{align}
    \chi^\text{sp}(\vec R = 0,\tau=0) &= \langle n \rangle - 2\langle n^{}_{\uparrow}n^{}_{\downarrow}\rangle, \label{eq:chi_sp_sumrule}\\
    \chi^\text{ch}(\vec R = 0,\tau=0) &= \langle n \rangle + 2\langle n^{}_{\uparrow}n^{}_{\downarrow}\rangle - \langle n \rangle^2,\label{eq:chi_ch_sumrule}
\end{align}
where $\chi^\text{sp/ch}$ is the spin/charge susceptibility, $\tau$ denotes imaginary time, and $n$ is the particle number operator.
While the susceptibilities $\chi^\text{sp/ch}$ are obtained from a Bethe-Salpeter equation from
the spin and charge vertices $\Gamma^\text{sp/ch}$ and the bare susceptibility $\chi^{0}=G^0\star G^0$
(and the non-interacting Green's function $G^0$),
the double occupation $\langle n^{}_{\uparrow}n^{}_{\downarrow}\rangle$
entering eqs.~\eqref{eq:chi_sp_sumrule} and \eqref{eq:chi_ch_sumrule} is \textit{a priori} unknown.
In the multi-orbital TPSC approach relate the spin vertex $\Gamma^\text{sp}$ to the double
occupation $\langle n^{}_{\uparrow}n^{}_{\downarrow}\rangle$ by means of the ansatz
\begin{align}
	\Gamma^\text{sp} = U \frac{\langle n^{}_{\uparrow}n^{}_{\downarrow}\rangle}{\langle n^{}_{\uparrow}\rangle \langle n^{}_{\downarrow}\rangle},\label{eq: Usp_ansatz_1orb}
\end{align}
which corresponds to a Hartree-Fock like decoupling as motivated in Ref.~\cite{Vilk1994}.
This relation is used to solve the local spin sum rule (eq.~\eqref{eq:chi_sp_sumrule}) for the double occupation.
This ansatz allows for a fully self-contained TPSC formulation, and it was demonstrated that the
self-consistently determined double occupations are in good agreement with Quantum Monte-Carlo results at temperatures above the renormalized classical regime. Though, for lower temperatures in the moderately correlated regime 
this ansatz was found to significantly underestimate the double occupations~\cite{VilkTremblay1997, LeBlanc2013, LeBlanc2015, Zantout2020}.

Another option is to consider the double occupation as an external parameter which
can be obtained by other many-body methods with higher precision, such as DMFT, which
is one of the main goals of this article.

\subsection{Multi-orbital TPSC}

Following the same line of arguments as in the single-orbital TPSC approach one can extend the formalism to multi-orbital Hubbard models.
A detailed derivation and discussion can be found in Ref.~\cite{Zantout2021}.
Here, we focus on the main equations to motivate the idea of combining TPSC with DMFT.

The Bethe-Salpeter equation evaluates to the following expressions for the spin and charge susceptibility,
\begin{align}
\chi^\text{sp/ch}(q,iq_n) &= 2\left( 1 \mp \chi^{0}(q, iq_n)\Gamma^\text{sp/ch} \right)^{-1} \chi^{0}(q,iq_n),
\end{align}
where $q$ is a reciprocal lattice vector, $q_n$ is the $n$-th bosonic Matsubara frequency, and $\chi^{0}=G^0\star G^0$ is the non-interacting irreducible susceptibility given as a convolution $(\star)$ of two non-interaction Green's functions $G^0$.
The Hartree-Fock-like decoupling which leads to eq.~\eqref{eq: Usp_ansatz_1orb} can be also applied to the multi-orbital case\footnote{Note that this ansatz breaks particle-hole symmetry which is restored by averaging this expression with the particle-hole transformed expression as explained in Ref.~\cite{Zantout2021}. In the following we use the particle-hole symmetrized expression.} which results in
\begin{align}
\Gamma^{\nk{sp}}_{\alpha\alpha\alpha\alpha} &= A_{\alpha}^{\sigma} \nonumber \\ 
\Gamma^{\nk{sp}}_{\alpha\alpha\beta\beta} 
&= B^{\uparrow,\downarrow}_{\alpha \beta} - B^{\downarrow,\downarrow}_{\alpha \beta}, \label{eq: Usp_ansatz}
\end{align}
where
\begin{align}
A^{\sigma}_{\alpha} &= U_{\alpha \alpha } \frac{\braket{n^{}_{\alpha \sigma} n^{}_{\alpha,-\sigma}}}{\braket{n^{}_{\alpha,\sigma}} \braket{n^{}_{\alpha,-\sigma}}},\\
B^{\sigma,\sigma}_{\alpha\beta} &= (U_{\alpha\beta} - J_{\alpha\beta} ) \frac{\braket{n^{}_{\alpha,\sigma} n^{}_{\beta,\sigma}}}{\braket{n^{}_{\alpha,\sigma}} \braket{n^{}_{\beta,\sigma}}}, ~~~ \alpha \neq \beta, \\
B^{\sigma,-\sigma}_{\alpha\beta} &= U_{\alpha\beta} \frac{\braket{n^{}_{\alpha,\sigma} n^{}_{\beta,-\sigma}}}{\braket{n^{}_{\alpha,\sigma}} \braket{n^{}_{\beta,-\sigma}}},~~~ \alpha \neq \beta.
\end{align}
These ansatz equations provide no expression to determine the spin-vertex elements
for the orbital combinations $\alpha\beta\alpha\beta$ and $\alpha\beta\beta\alpha$.
A possible solution is to assume a symmetric form of the spin vertex, namely
\begin{equation}
\Gamma^{\nk{sp}}_{ \alpha \beta \beta \alpha} = \Gamma^{\nk{sp}}_{ \alpha \beta \alpha \beta} = \Gamma^{\nk{sp}}_{ \alpha \alpha \beta \beta }, \label{eq: Usp_symmetrization}
\end{equation}
that is motivated from the bare $\Gamma^\text{sp,0}$ which obeys this symmetry in the longitudinal particle-hole channel~\cite{Zantout2021}.
The bare and dressed spin/charge vertices then enter the multi-orbital TPSC self-energy expression~\cite{Zantout2021}
\begin{align}
    \Sigma^{}_{\alpha \delta,\sigma} &= \left[\Gamma^{\nk{ch}} \chi^{ch} \Gamma^{\nk{ch},0 } + \Gamma^{\nk{sp}} \chi^{sp} \Gamma^{\nk{sp},0 } \right]_{\delta \bar{\beta} \alpha \bar{\gamma}}\star G^0_{ \bar{\gamma} \bar{\beta} }.     \label{eq: sigma_tpsc}
\end{align}
The ansatz equations~\eqref{eq: Usp_ansatz} and~\eqref{eq: Usp_symmetrization} together with the local spin and charge sum rules for the orbital index combinations for $\alpha\alpha\alpha\alpha$, $\alpha\alpha\beta\beta$, and $\alpha\beta\alpha\beta$ form a system of determined equations that can be solved to obtain the elements of the effective spin/charge vertices.

The shortcomings of the ansatz equations for the double occupation observed in the single-orbital case become visible when approaching the renormalized classical regime or at large interaction strength~\cite{VilkTremblay1997}.
These also seem to transfer to the multi-orbital case, where a qualitatively different
behavior as a function of interaction strength was found for the equal-spin double occupation $\langle n^{}_{\alpha,\sigma} n^{}_{\beta,\sigma} \rangle$ 
in TPSC compared to DMFT~\cite{Zantout2021}.
Furthermore, the TPSC double occupations can result in negative diverging charge vertices $\Gamma^\text{ch}_{\alpha\alpha\beta\beta}$, leading to unphysical negative spectral weight.
These unphysical solutions can be avoided by restricting the charge vertex $\Gamma^\text{ch}_{\alpha\alpha\beta\beta}$ to non-negative values, at the cost of violating the corresponding local charge sum rules~\cite{Zantout2021}.

These shortcomings of the ansatz equations for the double-occupations pose
a significant limitation of the multi-orbital formulation of TPSC,
especially in the moderately to stronger correlated regime.
In the following we will show how these limitations can be overcome by the combination 
of DMFT-derived double occupations with TPSC,
which we refer to as $\langle nn \rangle^{}_{\text{DMFT}}$-TPSC.

\subsection{$\langle nn \rangle^{}_{\text{DMFT}}$-TPSC}

In the multi-orbital formulation of TPSC there are not enough ansatz equations to 
determine the elements $\Gamma^{sp}_{\alpha\beta\alpha\beta}$ and $\Gamma^{sp}_{\alpha\beta\beta\alpha}$
and double-occupations self-consistently without further assumptions or approximations.
For this reason, we propose to use the double occupations obtained from a different many-body method
for the same model, such as DMFT. Together with the local spin and charge sum rules for the susceptibilities 
$\chi^{sp/ch}_{\alpha\alpha\alpha\alpha}$, $\chi^{sp/ch}_{\alpha\alpha\beta\beta}$ and $\chi^{sp/ch}_{\alpha\beta\alpha\beta}$
the corresponding vertex elements 
$\Gamma^{sp/ch}_{\alpha\alpha\alpha\alpha}$, $\Gamma^{sp/ch}_{\alpha\alpha\beta\beta}$ and $\Gamma^{sp/ch}_{\alpha\beta\alpha\beta}$ can be determined without the need for additional approximations or symmetry assumptions. 
Only for the remaining elements $\Gamma^{\nk{sp/ch}}_{ \alpha \beta \beta \alpha}$, for which
the sum rules do not result in density-density correlation terms, we retain the symmetry relation
\begin{equation}
    \Gamma^{\nk{sp/ch}}_{ \alpha \beta \beta \alpha} = \Gamma^{\nk{sp/ch}}_{ \alpha \beta \alpha \beta}. \label{eq: Usp_DMFT_symmetrization}
\end{equation}
Furthermore, this approach allows for restoring crossing symmetry in the multi-orbital TPSC formalism~\footnote{
The enforcement of crossing symmetry is in principle also in the pure multi-orbital TPSC approach possible but it was not done in Ref.~\cite{Zantout2019, Zantout2020, Zantout2021} as the symmetry $\Gamma^\text{sp,0}_{\alpha\alpha\beta\beta}=\Gamma^\text{sp,0}_{\alpha\beta\alpha\beta}$ is only valid in the longitudinal particle-hole channel and this relation is used to argue for the symmetrization in eq.~\eqref{eq: Usp_symmetrization}. 
Keeping the symmetrization in the spin vertex while including the transversal particle-hole channel leads to an inconsistency in the argumentation.
} 
by averaging the self-energy expressions from the longitudinal and transversal particle-hole channel~\cite{Tremblay2012, Zantout2020} which yields for the non-zero elements
of the spin vertex
\begin{align}
\Gamma^{\nk{sp,0}}_{\alpha \alpha \alpha \alpha} &= 3U_{\alpha\alpha}/2 \nonumber \\
\Gamma^{\nk{sp,0}}_{\alpha \alpha \beta  \beta } &= 2J_{\alpha\beta}  -U_{\alpha\beta}/2 \nonumber  \\
\Gamma^{\nk{sp,0}}_{\alpha \beta  \alpha \beta } &= J_{\alpha\beta}/2 +U_{\alpha\beta} \label{eq: Usp0_DMFT} \\
\Gamma^{\nk{sp,0}}_{\alpha \beta  \beta  \alpha} &= 3J_{\alpha\beta}/2 \nonumber ,
\end{align}
and charge vertex
\begin{align}
\Gamma^{\nk{ch,0}}_{\alpha \alpha \alpha \alpha} &= U_{\alpha\alpha}/2 \nonumber \\
\Gamma^{\nk{ch,0}}_{\alpha \alpha \beta  \beta } &= 3U_{\alpha\beta}/2-J_{\alpha\beta} \label{eq: Uch0_DMFT} \\
\Gamma^{\nk{ch,0}}_{\alpha \beta  \alpha \beta } &=
\Gamma^{\nk{ch,0}}_{\alpha \beta  \beta  \alpha} = J_{\alpha\beta}/2 \nonumber ,
\end{align}
for the vertices in the self-energy expression eq.~\eqref{eq: sigma_tpsc}.
This variant of the multi-orbital formulation of TPSC that 
determines the effective charge and spin vertices consistent with 
another many-body method we call $\langle nn \rangle^{}_{\text{X}}$-TPSC,
where X can in principle be any method that obtains accurate estimates
of the double occupations.

In this work we focus on double occupations that are extracted from DMFT 
which has been shown to obtain reliable estimates for local quantities
as long as non-local correlations are not too strong~\cite{Georges1996, Kotliar2004}.

\subsection{$\langle nn \rangle^{}_{\text{DMFT}}$-TPSC+$\Sigma^{}_\mathrm{DMFT}$}
As an additional step, we propose a scheme where the impurity self-energy
obtained from DMFT is combined with the non-local part of the TPSC self-energy,
which can be seen as an approximation to a fully self-consistent TPSC+DMFT
calculation. This scheme would correspond to a situation where the non-local correlation effects
are relevant, but not strong enough to have a significant feedback on the local correlation
effects obtained within DMFT. 
It improves upon the local and static interaction vertex included in TPSC,
which cannot consistently describe low and high energy features simultaneously due to
its static nature. For example, the vertices do not recover their bare values at large frequencies
as they are renormalized by low-energy spin- and charge-fluctuations.

In this scheme the effective momentum-dependent dynamical self-energy is given
as the local DMFT self-energy $\Sigma^\text{DMFT}$ combined with the non-local TPSC self-energy, namely
\begin{align}
&\Sigma(k, i^{}\omega^{}_n) \nonumber\\
=& \Sigma^{\nk{TPSC}}(k, i^{}\omega^{}_n) - \frac{1}{N}\sum_{k'}\Sigma^{\nk{TPSC}}(k', i^{}\omega^{}_n) + \Sigma^{\nk{DMFT}}(i^{}\omega^{}_n),\label{eq: Sigma_TPSCDMFT}  
\end{align}
where $k$ is a reciprocal lattice vector in the first Brillouin zone, $\omega^{}_n$ is the $n$-th fermionic Matsubara frequency, and $N$ the number of $k$-points. 
The idea of eq.~\eqref{eq: Sigma_TPSCDMFT} was previously developed for a different many-body approach~\cite{Schaefer2021}. 


\section{Results}
\label{sec: results}
We consider the two-orbital Hubbard model given by eq.~\eqref{Eq-Hubbard-H} to facilitate comparison with the multi-orbital TPSC formalism presented in Ref.~\cite{Zantout2021}.
The hopping terms are restricted to only nearest-neighbour hopping and to be orbital-diagonal $t_{\alpha\beta}=t_{\alpha\alpha}\delta_{\alpha\beta}$.
The only coupling between the orbitals is via the interaction terms given by $U$ and $J$.
We consider the half-filled case with one electron per orbital per site.
In the first part we will discuss the effect of the DMFT derived double occupations
on the TPSC two-particle quantities within $\langle nn \rangle^{}_{\text{DMFT}}$-TPSC.
These results apply in the same way to $\langle nn \rangle^{}_{\text{DMFT}}$-TPSC+$\Sigma^{}_\mathrm{DMFT}$,
as there is no feedback from the local DMFT self-energy on TPSC besides the inclusion of the DMFT double occupations.
The DMFT calculations are based on the ALPSCore continuous-time Quantum Monte-Carlo impurity solver
in the hybridization expansion~\cite{ALPSCore,cthyb_alps}.

\subsection{Double occupations}
\label{subsec: two_particle}
\begin{table}[h]
\begin{tabular}{| c | c | c || c | c || c | c |} 
 \hline
 $U/J=5$ & 
 \multicolumn{2}{c||}{$\langle n^{}_{\alpha,\uparrow}n^{}_{\alpha,\downarrow}\rangle$ } & 
 \multicolumn{2}{c||}{$\langle n^{}_{\alpha,\uparrow}n^{}_{\beta,\downarrow}\rangle$ } &
 \multicolumn{2}{c|}{$\langle n^{}_{\alpha,\uparrow}n^{}_{\beta,\uparrow}\rangle$ } \\
 $T/t$ & DMFT & TPSC & DMFT & TPSC & DMFT & TPSC  \\ [0.5ex] 
 \hline\hline
 0.5  & 0.1734 & 0.1571 & 0.2183 & 0.2361 & 0.2518 & 0.2692\\
 0.4  & 0.1759 & 0.1562 & 0.2193 & 0.2367 & 0.2517 & 0.2714\\
 0.3  & 0.1786 & 0.1541 & 0.2207 & 0.2376 & 0.2513 & 0.2759\\
 0.25 & 0.1796 & 0.1502 & 0.2205 & 0.2387 & 0.2507 & 0.2826\\
 \hline
\end{tabular}


\begin{tabular}{| c | c | c || c | c || c | c |} 
 \hline
 $U/J=3$ & 
 \multicolumn{2}{c||}{$\langle n^{}_{\alpha,\uparrow}n^{}_{\alpha,\downarrow}\rangle$ } & 
 \multicolumn{2}{c||}{$\langle n^{}_{\alpha,\uparrow}n^{}_{\beta,\downarrow}\rangle$ } &
 \multicolumn{2}{c|}{$\langle n^{}_{\alpha,\uparrow}n^{}_{\beta,\uparrow}\rangle$ } \\
 $T/t$ & DMFT & TPSC & DMFT & TPSC & DMFT & TPSC  \\ [0.5ex] 
 \hline\hline
 0.5  & 0.1595 & 0.1477 & 0.2197 & 0.213 & 0.2697 & 0.2832\\
 0.4  & 0.1614 & 0.1437 & 0.22204 & 0.213 & 0.2691 & 0.2899\\
 0.3  & 0.1646 & - & 0.2213 & - & 0.2685 & -\\
 0.25 & 0.1663 & - & 0.2220 & - & 0.2676 & -\\
 \hline
\end{tabular}
\caption{Double occupations as obtained from TPSC and DMFT at $U/t=3$ for different Hund's coupling $J$ and temperatures $T$.
Missing data points correspond to parameters where no TPSC convergence could be achieved.
}
\label{tab: double occ}
\end{table}
\begin{figure*}[t]
    \includegraphics[width=\linewidth]{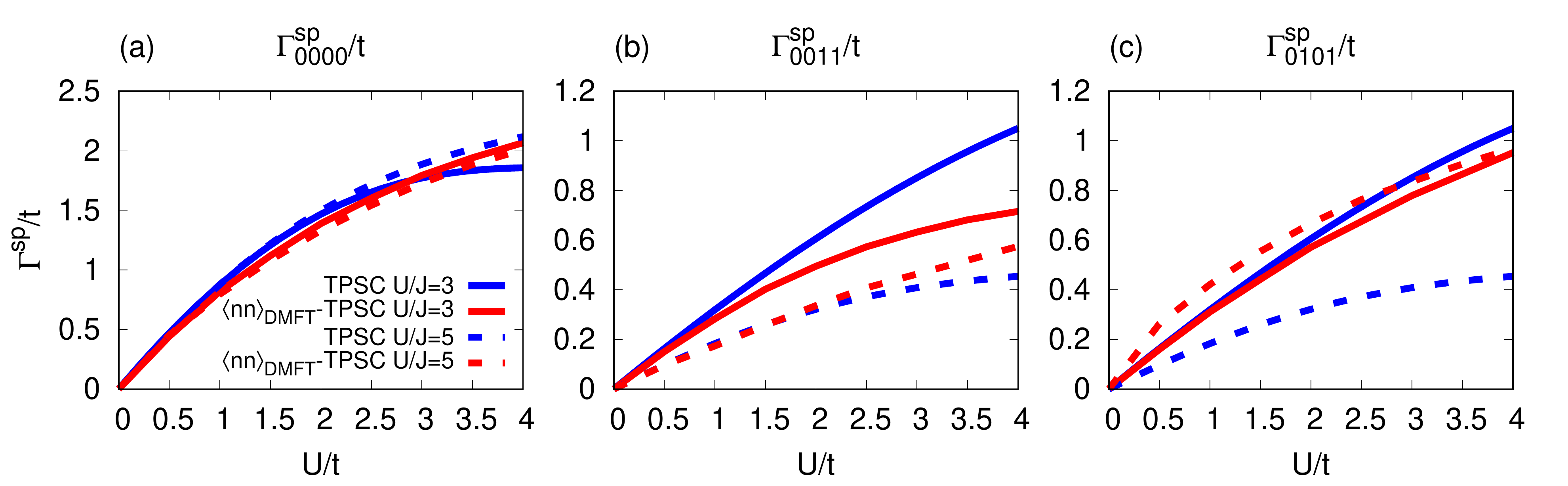}
    \caption{Spin vertex components (a) $\Gamma^{sp}_{\alpha\alpha\alpha\alpha}$, (b) $\Gamma^{sp}_{\alpha\alpha\beta\beta}$, and (c) $\Gamma^{sp}_{\alpha\beta\alpha\beta}$ as a function of $U/t$ for $U/J=3$ (full lines) and $U/J=5$ (dashed lines), within TPSC and $\braket{nn}^{}_{\text{DMFT}}$-TPSC. While both methods show a similar dependence on $U$, we find 
    a nontrivial effect of the DMFT double occupations when included in TPSC: The off-diagonal vertex elements
    are increased for small Hund's coupling $J$, but are reduced for large $J$, with significant modification of the
    $\Gamma^\text{sp}_{\alpha\beta\alpha\beta}$ element. This element is determined by its own sum rule in $\braket{nn}^{}_{\text{DMFT}}$-TPSC, in contrast to TPSC, where it is set equal to $\Gamma^\text{sp}_{\alpha\alpha\beta\beta}$.
    The remaining non-zero element $\Gamma^{sp}_{\alpha\beta\beta\alpha}$ is equal to $\Gamma^{sp}_{\alpha\beta\alpha\beta}$. 
    TPSC data taken from Ref.~\cite{Zantout2021}.
    }
    \label{fig: Usp_TPSC_vs_DMFT+TPSC}
\end{figure*}
In Table~\ref{tab: double occ} we show the double occupations obtained from multi-orbital TPSC compared to DMFT,
for $U/t=3$ for moderate ($J=U/5$) and large ($J=U/3$) values of the Hund's coupling at different temperatures $T/t$.
We find the general trend that the TPSC equal-orbital double occupations are about $5-15\%$ smaller than the ones obtained from DMFT indicating stronger local correlation effects in TPSC similar to the single-orbital case~\cite{Schaefer2021}.
Furthermore, TPSC shows a monotonous increase as a function of temperature as it does not capture the reduction of the double occupation induced by favoring localization to increase spin entropy~\cite{LeBlanc2015,Schaefer2021,Sushchyev2022}.
We attribute this behavior in TPSC to the mean-field like ansatz equation~\eqref{eq: Usp_ansatz}.
While this effect is observed in TPSC for weakly correlated systems~\cite{Schaefer2021}, it is not captured in strongly correlated systems~\cite{Vilk1997} such as in this case.
While DMFT obtains this reduction qualitatively, it is known to overestimate this effect~\cite{Dare2007}.
For the same reason we also observe a reversed temperature trend between TPSC and DMFT for $\langle n^{}_{\alpha,\uparrow}n^{}_{\beta,\uparrow}\rangle$.
The remaining double occupations follow qualitatively the same temperature dependence with TPSC double occupations being in general 5-10\% larger. Compared to moderate values of the Hund's coupling $U/J=5$ we observe for larger
values $U/J=3$ the same trends, but the root search for determining the values of the 
spin vertex becomes increasingly unstable, which did not allow us to obtain converged results for 
temperatures below $T<0.35$. Especially in such cases a more reliable way of obtaining the double occupations
as input for the TPSC calculation such as from a DMFT calculation is promising, and will be explored in the next section.

%
\subsection{Spin and charge vertices}
As the effective local vertices in TPSC are determined by imposing local sum rules that depend on
the double occupations, the DMFT derived double occupations have a direct influence on the effective
interaction vertices $\Gamma^{sp/ch}$ in $\langle nn \rangle^{}_{\text{DMFT}}$-TPSC.
In Fig.~\ref{fig: Usp_TPSC_vs_DMFT+TPSC} we show the spin vertex $\Gamma^\text{sp}$ as a function of interaction strength $U/t$ for different values of $U/J$.
In contrast to all the intra- and inter-orbital double occupations, which show substantial differences between TPSC and DMFT,
we observe a selective effect on the different elements of the spin-vertex.
Overall the same functional dependence of $\Gamma^\text{sp}$ on $U/t$ as in TPSC is retained,
with Kanamori-Brueckner screening at larger interaction values. 
While the diagonal elements $\Gamma^{sp}_{\alpha\alpha\alpha\alpha}$ stay almost unchanged,
the off-diagonal elements differ significantly and the changes are sensitive to the Hund's coupling $J$.
For small $J$ the inclusion of the DMFT double occupation leads to an increase of the effective spin vertex element,
while for large $J$ they are reduced compared to the TPSC value. The increase is especially pronounced in 
the $\alpha\beta\alpha\beta$ element for small $J$. We attribute this to the mean-field like decoupling in the 
TPSC ansatz, which seems to perform better for larger values of $J$ where on average the bare interaction
elements are reduced.
Most importantly, we observe that the $\Gamma^\text{sp}_{\alpha\beta\alpha\beta}$ element, which 
in $\langle nn \rangle^{}_{\text{DMFT}}$-TPSC is now determined by its own sum rule,
can differ up to a factor of $2$ from the $\Gamma^{\nk{sp}}_{ \alpha \alpha \beta \beta }$ element.
This indicates that the ansatz $\Gamma^\text{sp}_{\alpha\beta\alpha\beta} = \Gamma^{\nk{sp}}_{ \alpha \alpha \beta \beta }$
in multi-orbital TPSC due to a lack of sum rules constitutes a significant approximation and is not able to capture the renormalization of the off-diagonal spin vertex elements.

\begin{figure*}[t]
    \centering
    \includegraphics[width=\linewidth]{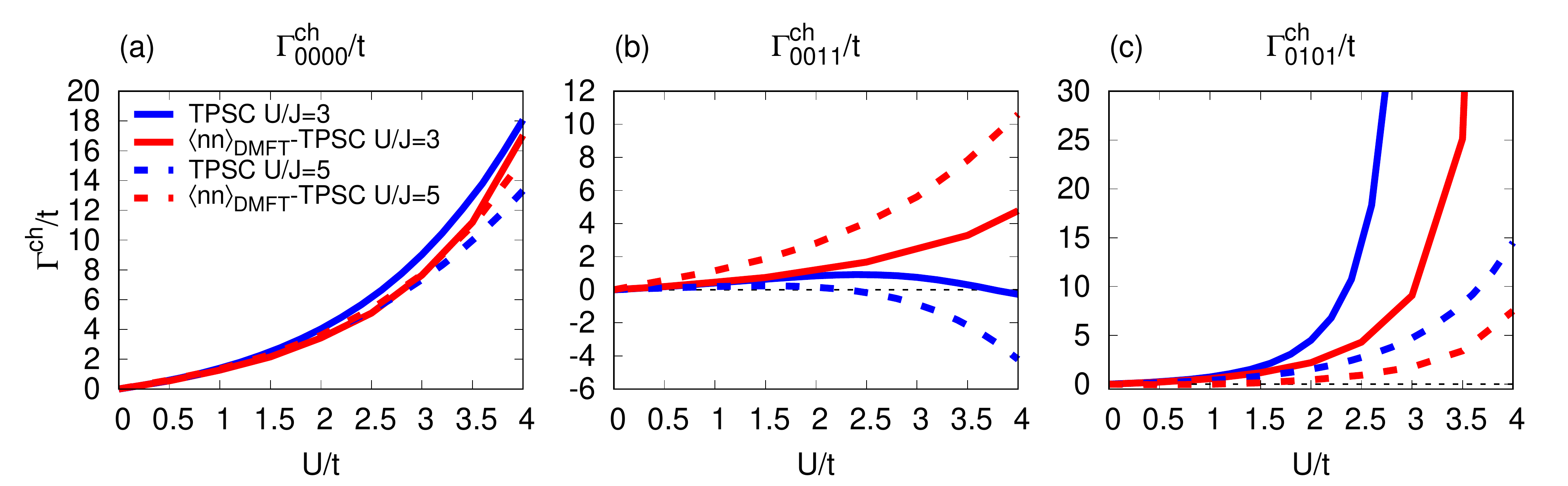}
    \caption{Charge vertex components (a) $\Gamma^{ch}_{\alpha\alpha\alpha\alpha}$, (b) $\Gamma^{ch}_{\alpha\alpha\beta\beta}$, and (c) $\Gamma^{ch}_{\alpha\beta\alpha\beta}$ as a function of $U/t$ for $U/J=3$ (dashed lines) and $U/J=5$ (full lines), within TPSC (blue) and $\braket{nn}^{}_{\text{DMFT}}$-TPSC (red). 
    While the inclusion of the DMFT double occupations only leads to a small increase in the diagonal vertex elements,
    the $\Gamma^\text{ch}_{\alpha\alpha\beta\beta}$ elements, which in TPSC are negative and lead to unphysical solutions, become strictly positive.
    The $\Gamma^\text{ch}_{\alpha\beta\alpha\beta}$ elements are reduced significantly in $\braket{nn}^{}_{\text{DMFT}}$-TPSC,
    with negligible negative values at small interactions.
    TPSC results are taken from Ref.~\cite{Zantout2021}.
    }
    \label{fig: U_vs_Uch}
\end{figure*}
In Fig.~\ref{fig: U_vs_Uch} we show the charge vertex $\Gamma^\text{ch}$ obtained from TPSC and $\braket{nn}^{}_\text{DMFT}$-TPSC.
While the diagonal elements $\Gamma^\text{ch}_{\alpha\alpha\alpha\alpha}$ (Fig.~\ref{fig: U_vs_Uch} (a)) monotonously increase as a function of $U$ and show only minor differences between the two approaches,
we find the most significant improvement in the $\Gamma^\text{ch}_{\alpha\alpha\beta\beta}$ elements (Fig.~\ref{fig: U_vs_Uch} (b)):
The large negative values for the charge vertices $\Gamma^\text{ch}_{\alpha\alpha\beta\beta}$ observed in TPSC are absent in $\braket{nn}^{}_\text{DMFT}$-TPSC,
which resolves the problem of unphysical negative spectral weight contributions to the self-energy at larger values of $U/t$~\cite{Zantout2021}.
The $\Gamma^\text{ch}_{\alpha\beta\alpha\beta}$ element (Fig.~\ref{fig: U_vs_Uch} (c))
on the other hand is reduced compared to TPSC, which is likely due to a previous compensation
effect with the smaller or negative $\Gamma^\text{ch}_{\alpha\alpha\beta\beta}$ in order to fulfill the 
corresponding sum rule, which required larger values of $\Gamma^\text{ch}_{\alpha\beta\alpha\beta}$.
Still, we observe small negative values of $\Gamma^\text{ch}_{\alpha\beta\alpha\beta}$ at weak interaction,
but they are negligible due to their small relative magnitude $|\Gamma^\text{ch}_{\alpha\beta\alpha\beta}|/U\approx 10^{-2}$.

These results suggest that the DMFT double occupations can provide a significant improvement
over the multi-orbital TPSC ansatz equations in the multi-orbital case, as they 
result in physical vertices, in contrast to the multi-orbital TPSC approach.
While in general we cannot exclude the existence of certain scenarios where the vertex elements
can become negative and large (as likely might be the case where the approximation of a static vertex is not appropriate),
we expect this to be a general result for many systems of interest,
and that $\langle nn \rangle^{}_{\text{DMFT}}$-TPSC provides access to more strongly correlated systems that were previously out of reach within multi-orbital TPSC.


\subsection{Antiferromagnetic spin fluctuations}
\begin{figure}[t]
    \includegraphics[width=\linewidth]{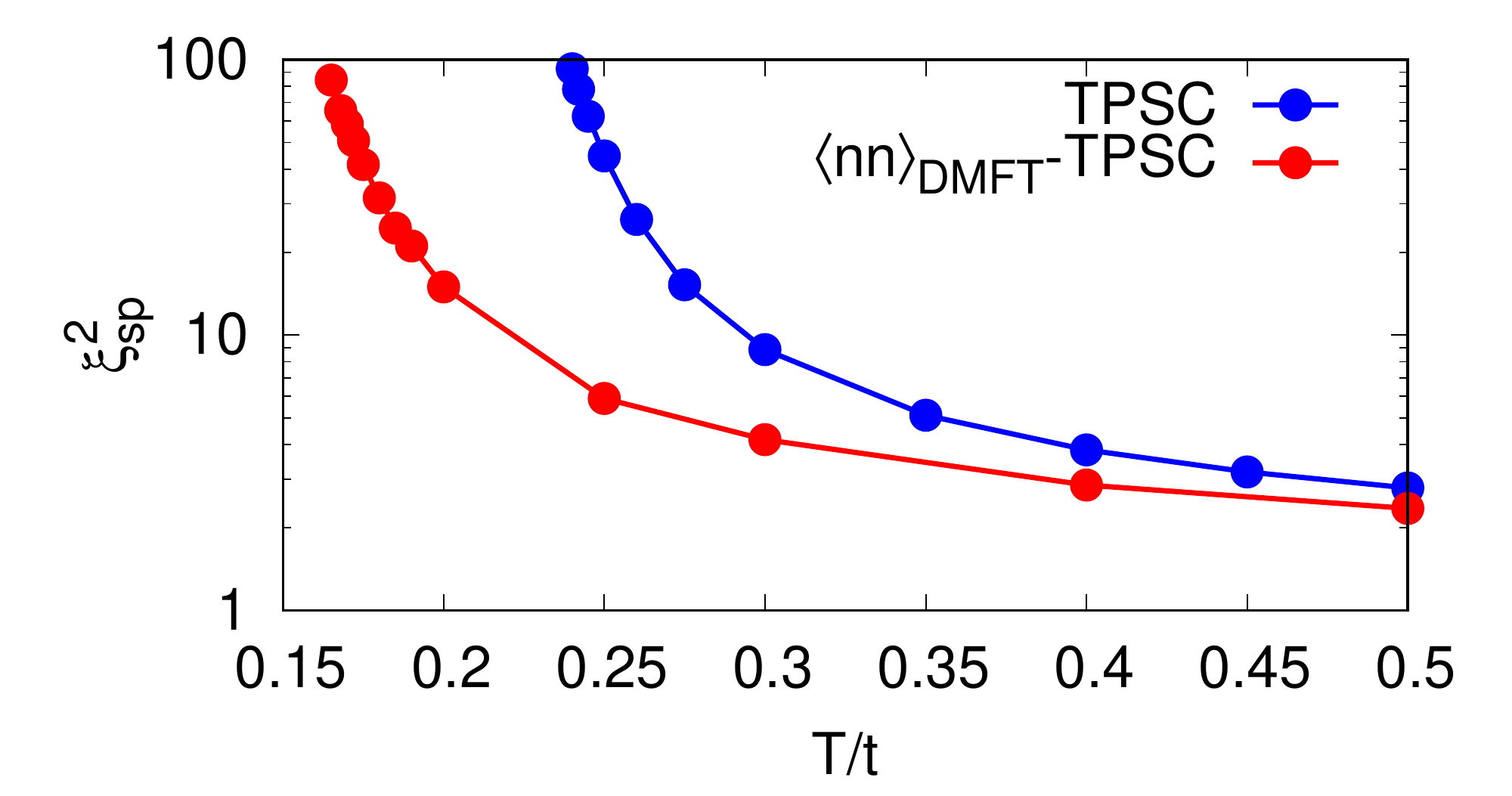}
    \caption{Antiferromagnetic spin correlation length $\xi^{2}_\text{sp}$ as a function of  temperature $T/t$ at $U/t=3, U/J=5$. While TPSC fulfills the Mermin-Wagner theorem with the divergence at $T=0$, it overestimates the spin correlation length. $\braket{nn}^{}_{\text{DMFT}}$-TPSC
    obtains a reduced correlation length for all temperatures considered.
    }
    \label{fig: xi_T}
\end{figure}
\begin{figure*}[t]
    \centering
    \includegraphics[width=\linewidth]{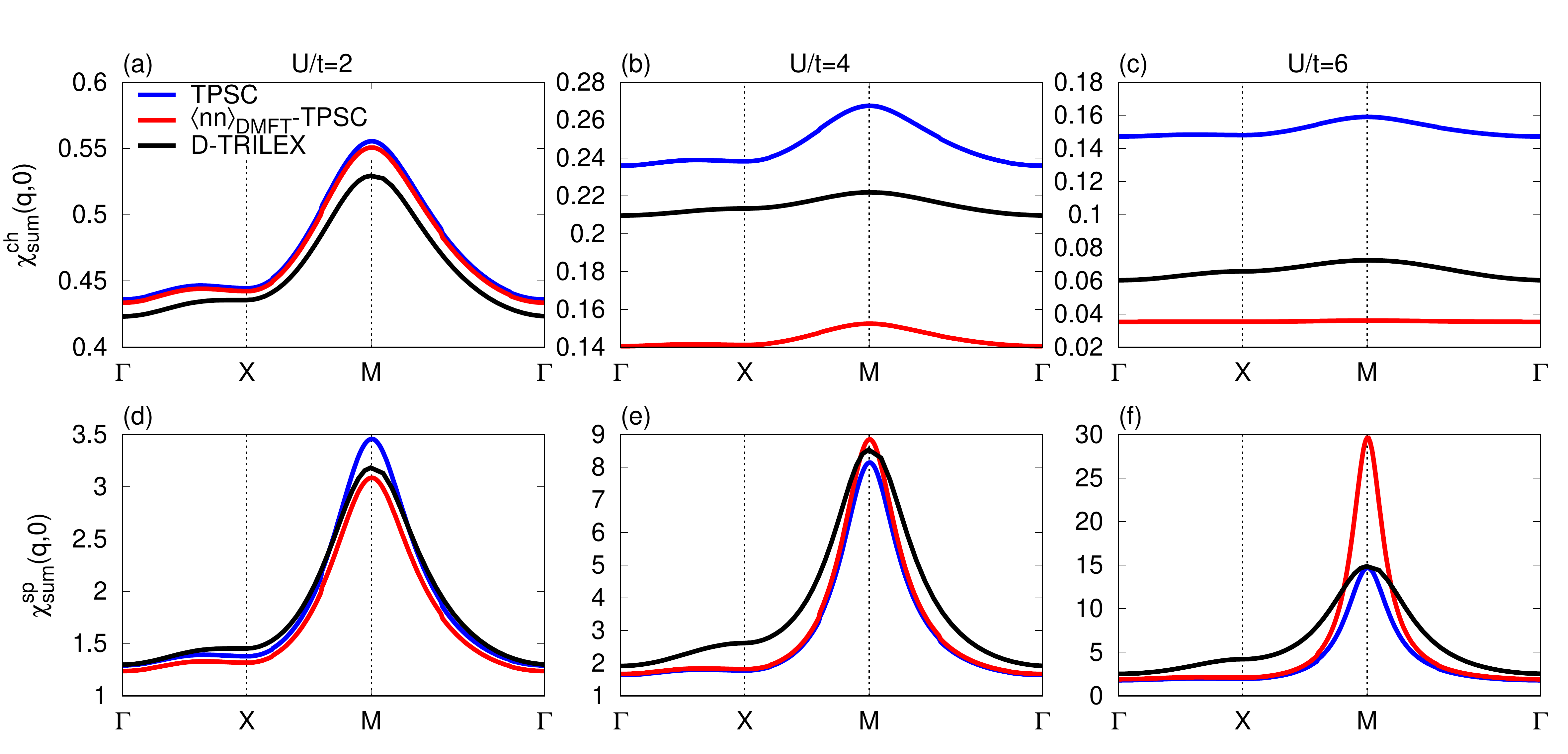}
    \caption{Summed charge (upper panels) and spin (lower panels) susceptibilities $\sum_{\alpha, \beta}\chi^{sp/ch}_{\alpha\alpha\beta\beta}(q,0)$ along $\Gamma-X-M-\Gamma$ for 
    (a), (d) $U/t=2$, (b), (e) $U/t=4$, (c),(f) $U/t=6$ and $U/J=4$ obtained from TPSC, $\langle nn \rangle^{}_{\text{DMFT}}$-TPSC, and D-TRILEX. In both cases we observe the suppression/enhancement of charge/spin fluctuations with increasing $U/t$. While TPSC underestimates this trend,
    inclusion of the DMFT double occupations within $\langle nn \rangle^{}_{\text{DMFT}}$-TPSC+$\Sigma_{DMFT}$
    lead to an improved agreement with D-TRILEX at moderate interaction values.
    For strong interactions the suppression/enhancement of charge/spin fluctuations is overestimated
    in $\langle nn \rangle^{}_{\text{DMFT}}$-TPSC+$\Sigma_{DMFT}$. D-TRILEX results are taken from Ref.~\cite{Vandelli2022}.
    }
    \label{fig: suscep_DTrilex}
\end{figure*}
In order to assess the effect of the DMFT-derived double occupations on
the spin fluctuations in TPSC we define the antiferromagnetic spin correlation length as the ratio of
the spin- and bare susceptibility at the $M$ point
\begin{equation}
    \xi^{2}_\text{sp} := \frac{\chi^\text{sp}_{\alpha\alpha\alpha\alpha}(\vec q= (\pi,\pi),\omega=0)}{\chi^0_{\alpha\alpha\alpha\alpha}(\vec q= (\pi,\pi),\omega=0)}. \label{eq: spin_correlation_length}
\end{equation}
Fig.~\ref{fig: xi_T} shows $\xi^{2}_\text{sp}$ at $U/t=3$ and $U/J=5$,
as a function of temperature.
The spin correlation length increases upon lowering the temperature, representing the increasing antiferromagnetic
fluctuations in the two-dimensional Hubbard model~\cite{Vilk1994,Schaefer2021},
but only diverges at $T=0$ as TPSC obeys the Mermin-Wagner theorem.
We observe that the spin correlation length in $\langle nn \rangle^{}_{\text{DMFT}}$-TPSC 
is smaller than in multi-orbital TPSC at the same temperature, up to more than an order of magnitude
at lower temperatures. This is indicative for the overestimation of the strength
of spin fluctuations in TPSC~\cite{Tremblay2012,Schaefer2021}, which is significantly improved
when deriving the spin vertex from the DMFT double occupations (see Table~\ref{tab: double occ}).
In fact, $\langle nn \rangle^{}_{\text{DMFT}}$-TPSC can also be seen as an effective
way of mimicking frequency-dependent vertex corrections in TPSC via the double occupations.
A similar improvement can be seen in TPSC+~\cite{TPSCplus,Schaefer2021}, which includes effective dynamical
vertex corrections by a feedback of the self-energy into the propagators.
We note that this behavior is dependent on the model parameters, as we find that for a given temperature
$\langle nn \rangle^{}_{\text{DMFT}}$-TPSC can lead both to a reduction of the antiferromagnetic  correlations
at small values of $U$ and also to an enhancement at larger values of $U$, as will be shown in the next section.

\subsection{Susceptibilities}
\label{sec: suscep}
In the following sections we will perform a benchmark of the $\langle nn \rangle^{}_{\text{DMFT}}$-TPSC
and $\langle nn \rangle^{}_{\text{DMFT}}$-TPSC$+\Sigma_{DMFT}$ approach on a two-orbital
Hubbard model and compare our results to the D-TRILEX approach~\cite{Stepanov2019}.  
D-TRILEX is an approximation to the dual boson method
and has recently been extended to multi-orbital systems~\cite{Vandelli2022}.
It treats charge and spin fluctuations on the same footing, and by the inclusion of the dynamical but local three-point vertex
is able to describe non-local correlation and screening effects in strongly interacting systems.
This makes it a reasonable reference method for our improved TPSC schemes, which work
with a simplified static and local two-point vertex.

\begin{figure*}
    \centering
    \includegraphics[width=\linewidth]{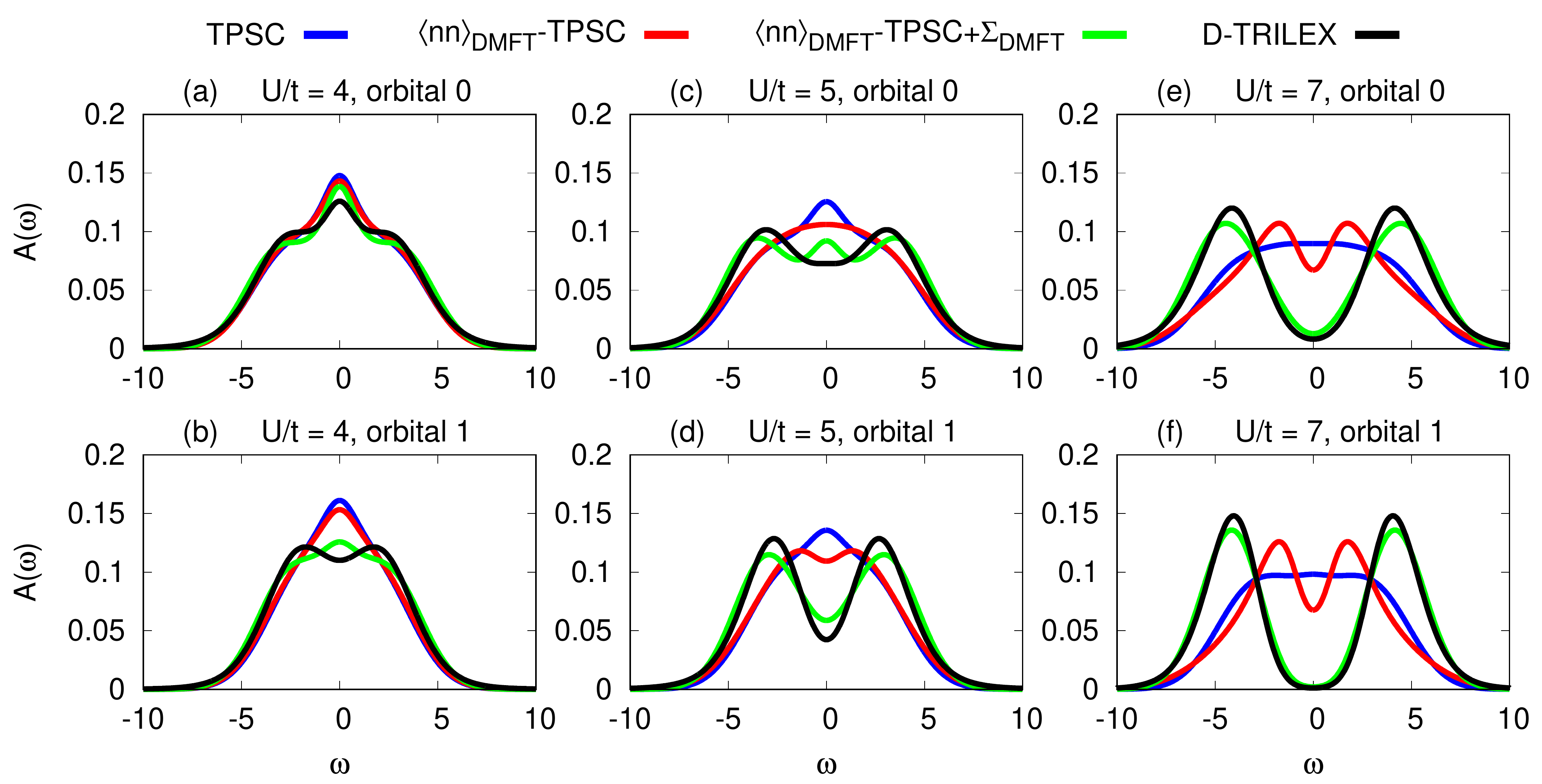}
    \caption{The local spectral function $A(\omega)$ obtained for the two-orbital Hubbard model (see main text) from TPSC (blue), $\langle nn \rangle^{}_{\text{DMFT}}$-TPSC (red), $\langle nn \rangle^{}_{\text{DMFT}}$-TPSC+$\Sigma_{DMFT}$ (green), and D-TRILEX (black) for interaction values (a),(b) $U/t=4$, (c),(d) $U/t=5$, and (e), (f) $U/t=7$. TPSC underestimates the correlation strength for all parameters considered, and is not able to obtain the reduction of spectral weight at the Fermi level for stronger interactions.
    Incorporating the DMFT double occupations in $\langle nn \rangle^{}_{\text{DMFT}}$-TPSC provides a 
    notable improvement over TPSC, while only $\langle nn \rangle^{}_{\text{DMFT}}$-TPSC+$\Sigma_{DMFT}$ is 
    able to access the Mott insulating phase in the second orbital, and obtains a qualitative agreement
    with the D-TRILEX result.   D-TRILEX results are taken from Ref.~\cite{Vandelli2022}.
    }
    \label{fig: spectral_function}
\end{figure*}

In the following we consider the half-filled two-orbital Hubbard square lattice at temperature $T/t=0.5$ with an orbital-dependent
nearest-neighbour hopping
\begin{align}
    t_{00}=1.0,\ t_{11}=0.75 \label{eq: hopping_DTRILEX}
\end{align}
where the bandwidth of the second orbital is reduced by $25\%$.
In Fig.~\ref{fig: suscep_DTrilex} we show the momentum-resolved summed spin and charge susceptibilities
\begin{align}
    \chi^{ch/sp}_\text{sum}(q,0) &:= \sum_{\alpha,\beta} \chi^{ch/sp}_{\alpha\alpha\beta\beta}(q,0),
\end{align}
for $U/t=2,4,6$ and $U/J=4$ obtained within multi-orbital TPSC and $\langle nn \rangle^{}_{\text{DMFT}}$-TPSC, and 
compare them to the D-TRILEX results from Ref.~\cite{Vandelli2022}.
In general we find good qualitative and quantitative agreement between the three methods for small interactions
in both the charge- and spin susceptibility (Fig.~\ref{fig: suscep_DTrilex}(a) and (d)).
They exhibit pronounced peaks at the $M=(\pi,\pi)$ point, corresponding to spin- and charge fluctuations
with wave vector $\vec{q}=(\pi,\pi)$, with the dominating spin fluctuations indicating antiferromagnetic
order as the main instability of the system. All methods show a reduction of the charge susceptibility and  increase
of the spin susceptibility for increasing interaction $U/t=2..6$.
While the differences at $U/t=2$ between TPSC and $\langle nn \rangle^{}_{\text{DMFT}}$-TPSC are
small, one clearly observes a shift towards the D-TRILEX result when using the DMFT derived double occupations,
bringing the spin susceptibility of $\langle nn \rangle^{}_{\text{DMFT}}$-TPSC
in almost  perfect agreement with D-TRILEX.
At stronger interactions we observe that multi-orbital TPSC significantly overestimates the charge susceptibility, but
underestimates the spin susceptibility, and increasingly starts to deviate from the D-TRILEX result.
On the other hand, $\langle nn \rangle^{}_{\text{DMFT}}$-TPSC always shows the tendency to correct 
the difference to D-TRILEX, but overestimates the reduction of the charge susceptibility 
and the enhancement of the spin susceptibility for larger interactions
(Fig.~\ref{fig: suscep_DTrilex}(b), (c), and (f)).
While multi-orbital TPSC is not able to capture the Mott transition in either orbital and also
obtains a charge susceptibility that is significantly too large at $U/t=6$, 
the DMFT double occupations effectively encode 
the insulating nature of the system in $\langle nn \rangle^{}_{\text{DMFT}}$-TPSC,
which shows an almost vanishing charge susceptibility at this interaction value.

\subsection{Spectral function}

So far, we have only discussed quantities such as the double occupation, effective vertices or susceptibilities
within $\langle nn \rangle^{}_{\text{DMFT}}$-TPSC. As these quantities only depend on the non-interacting Green's function, bare interaction and DMFT double occupations,
they are not affected by a replacement of the local TPSC self-energy with the DMFT impurity self-energy 
as proposed for $\langle nn \rangle^{}_{\text{DMFT}}$-TPSC$+\Sigma_{DMFT}$.
On the other hand, the single-particle local spectral function
\begin{align}
    A(\omega) = \frac{-1}{\pi N}\sum_k \Im \left[ \omega + i0^+ - H_0(k) - \Sigma(k,\omega) \right]^{-1}
\end{align}
will be affected by the replacement, as it directly depends on the final self-energy. 
For the same two-orbital Hubbard model as in the previous section we compare
the two different methods with TPSC and D-TRILEX~\cite{Vandelli2022} at different interactions for $U/J=4$
in Fig.~\ref{fig: spectral_function}. 
Analytic continuation from the imaginary to the real frequency axis has been performed
by using the maximum entropy formalism~\cite{Maxent,ALPSCore,levi_maxent}.
In agreement with D-TRILEX we find the orbital with the larger bandwidth to be less correlated
than the one with the narrow bandwidth (see Figs.~\ref{fig: spectral_function}), and we observe an overall increase in correlation effects
as the interaction is increased.
All TPSC related methods differ considerably from each other: 
We obtain that multi-orbital TPSC underestimates the correlation strength the most and thus shows the largest
difference to the D-TRILEX results for all interaction values in both orbitals.
The approach is also not able to capture the large reduction of spectral weight at the Fermi level in the first orbital and 
a Mott-transition in the second orbital  at $U/t=7$ (Fig.~\ref{fig: spectral_function} (e) and (f)).
$\langle nn \rangle^{}_{\text{DMFT}}$-TPSC on the other hand, provides a considerable improvement over TPSC,
capturing the reduction of spectral weight qualitatively, but still underestimates the correlation strength.
The best agreement with the D-TRILEX benchmark is found for $\langle nn \rangle^{}_{\text{DMFT}}$-TPSC$+\Sigma_{DMFT}$,
which is able to capture the Mott transition and shows close agreement for all parameters considered,
albeit a remaining underestimation of the correlation strength.
These results show that the incorporation of more accurate double occupations
within the $\langle nn \rangle^{}_{\text{DMFT}}$-TPSC approach already leads to a notable improvement
over TPSC. Still, for strongly correlated systems the static TPSC vertex entering the self-energy remains a major
limitation as expected. This limitation can be drastically improved by a combination with the DMFT impurity self-energy
in $\langle nn \rangle^{}_{\text{DMFT}}$-TPSC$+\Sigma_{DMFT}$,
providing access to the Mott insulating phase which was previously not accessible in TPSC.


\section{Conclusion}
\label{sec: conclusion}

In this work we have presented two extensions of the multi-orbital TPSC approach~\cite{Zantout2021} that
are based on incorporating local quantities which can be obtained with higher precision
from a DMFT calculation in the TPSC formalism. 
The first extension, called $\langle nn \rangle^{}_{\text{DMFT}}$-TPSC,
consists in replacing the usual ansatz equations for the double occupations in TPSC
by the double occupations sampled in a DMFT calculation for the same system.
As the TPSC ansatz is based on a Hartree-Fock-like decoupling, 
this avoids additional approximations in the determination of the double occupations,
which are needed for the determination of the effective spin and charge vertices in TPSC.
We found this approach to be highly successful specifically for the multi-orbital
form of TPSC, as it removed the negative divergences of the charge vertex $\Gamma^{ch}$ observed
in the multi-orbital TPSC approach~\cite{Zantout2021}. Additionally, certain inter-orbital elements of the effective spin vertex,
which previously were determined by symmetries only valid for the bare interaction vertex,
can be determined explicitly, and indeed show an expected deviation from the bare interaction case.
This extension also provides access to lower temperatures that were previously 
inaccessible in TPSC, as divergences in the spin vertex are shifted to lower
temperatures.
Furthermore, it allows for the inclusion of the transversal particle-hole channel 
which restores crossing symmetry in the vertex functions.
Nevertheless, we note that the double occupations are influenced by non-local correlations which are not taken into account in DMFT, in particular in regions where strong order parameter fluctuations prevail~\cite{Rohringer2016, Stobbe2022}. 
See appendix~\ref{app: double occ} for further discussion on the double occupations and~\ref{app: internal_check} for a discussion on the internal consistency between single- and two-particle objects.

In the second proposed extension we replace the local part of the the TPSC self-energy by
the impurity self-energy of DMFT, called $\langle nn \rangle^{}_{\text{DMFT}}$-TPSC+$\Sigma_{DMFT}$.
This approach improves upon the static vertex included in TPSC by effectively incorporating 
a dynamical DMFT vertex in the local self-energy, while retaining non-trivial 
momentum-dependent correlation effects from TPSC at low computational costs.
We found this approach to provide further improved agreement with other many-body methods,
especially for local one-particle quantities such as the local spectral function.
This approach also extends the applicability of TPSC to systems with strong correlations,
as it is able to access the Mott-insulating phase, previously inaccessible in multi-orbital TPSC.
We note here that by replacing the local part of the self-energy, we still have contributions from non-local correlation effects.
From the diagrammatic point of view our quantity is therefore only semi-local.
An alternative approach can be constructed by means of the local Dyson equation, where the local Green's function is used to obtain the local self-energy, which allow us to subtract all non-local correlation effects but on the other hand does not yield the exact DMFT result in the limit of infinite connectivity.

These results show the potential of combining TPSC with DMFT for both local and non-local
quantities, and opens up the door towards further developments such as fully self-consistent TPSC+DMFT calculations and further applications to real materials.

\acknowledgments{
We would like to thank Andr\'e-Marie Tremblay, Chlo\'{e}-Aminata Gauvin-Ndiaye, Nicolas Martin and
Olivier Gingras for useful discussions. We acknowledge support from the Deutsche Forschungsgemeinschaft (DFG, German Research Foundation) through TRR 288-422213477 (Project B05) (A.R., R.V.) and through FOR 5249-449872909 (Project P4) (R.V.). 
}

\appendix
\section{Double occupations}
\label{app: double occ}
\begin{figure}[t]
    \centering
    \includegraphics[width=1\linewidth]{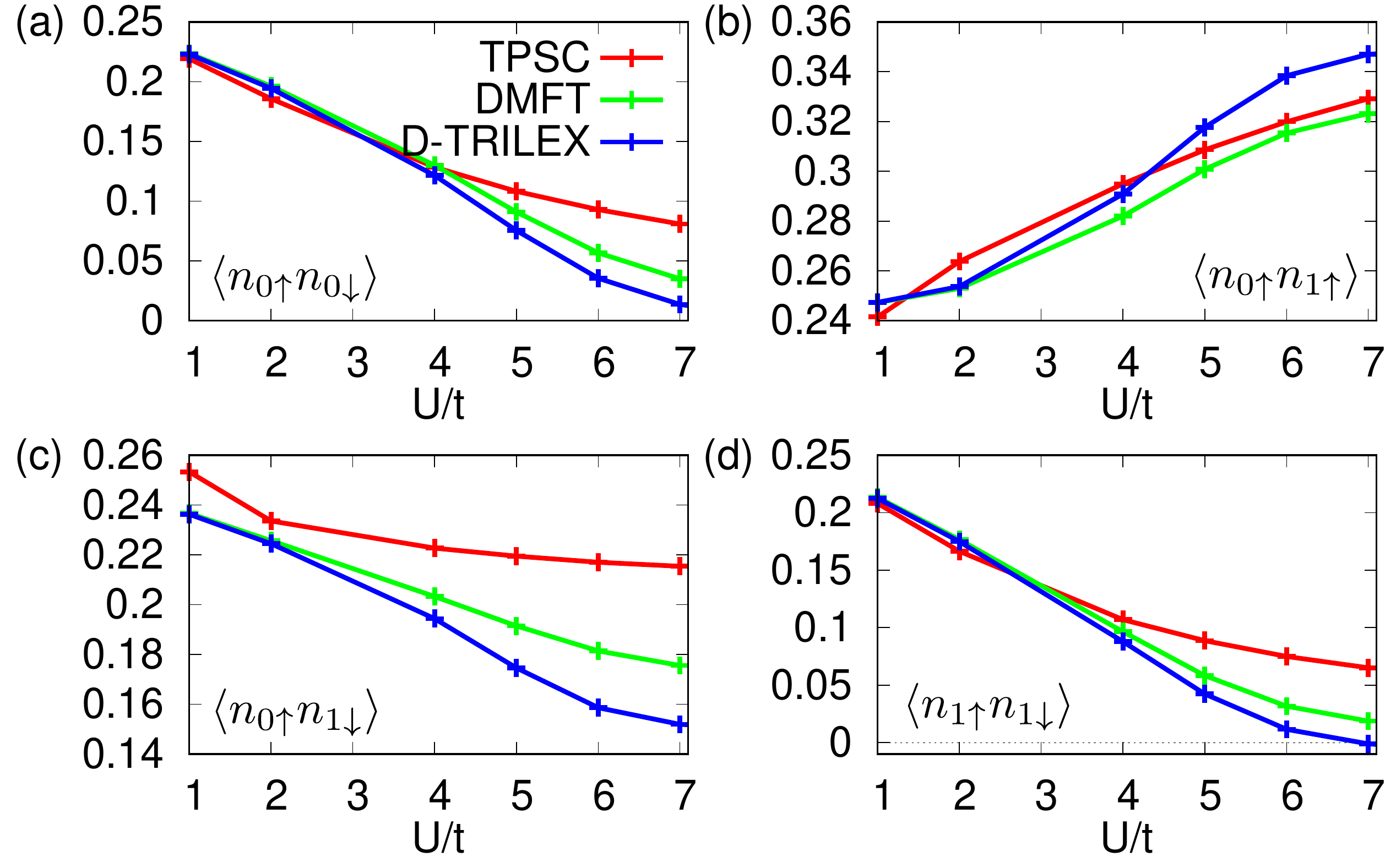}
    \caption{Comparison of the double occupations for the model described in sec.~\ref{sec: suscep}. We observe a small difference between all methods at low interaction strength ($U<3t$), but larger differences at higher interaction strength where nonlocal correlations become significant.
    In general the DMFT derived double occupations show a closer agreement with the D-TRILEX result than TPSC, except for the high-spin inter-orbital configuration, indicating that the DMFT result can serve as a possible improved starting point for the determination of the effective TPSC vertices.
        The results for the D-TRILEX double occupancies have been provided by the 
        authors of Ref. \cite{Vandelli2022}.
    }
    \label{fig: double_occ}
\end{figure}

In order to compare the resulting values for the double occupations for the different approaches and the influence of nonlocal correlations, we show the double occupations from TPSC, DMFT, and D-TRILEX in Fig.~\ref{fig: double_occ}.
As discussed in the main text the deviation between the TPSC and DMFT double occupations is small at weak interaction strength ($U/t<3$) but becomes larger for stronger interactions, as TPSC is not able to capture the Mott insulator phase.
In general the deviation between the DMFT and D-TRILEX double occupations is smaller, indicating that indeed the DMFT double occupations are a better starting point than the TPSC derived ones. When nonlocal correlations become strong at larger interaction values, the DMFT and D-TRILEX result also start to differ, with the general trend that D-TRILEX obtains a larger correlation strength, i.e. reduced low-spin double occupations and enhanced high-spin configurations.

\section{Internal consistency check}
\label{app: internal_check}

In the original TPSC formulation~\cite{VilkTremblay1997} the $\mathrm{tr}(\Sigma G)$ sum rule,
\begin{align}
&\mathrm{tr}(\Sigma G)^{}_{\beta,\sigma}\nonumber\\
&=\sum_{\alpha}U_{\alpha\beta}\langle n_{\alpha,-\sigma}n_{\beta,\sigma}\rangle +\sum_{\substack{\alpha\\\alpha\neq \beta}}(U_{\alpha\beta}-J_{\alpha\beta})\langle n_{\alpha,\sigma}n_{\beta,\sigma}\rangle\nonumber\\
&~-\sum_{\substack{\alpha\\\alpha\neq \beta}}J_{\alpha\beta}(\langle n_{\alpha,\sigma}n_{\beta,\sigma}\rangle-\langle n_{\alpha,-\sigma}n_{\beta,\sigma}\rangle)\nonumber\\
&~-\frac{1}{\tilde\beta N_{\vec{q}}}\sum_{\substack{q,\alpha\\\alpha\neq \beta}}\frac{J_{\alpha\beta}}{2}\left(\chi^{sp}_{\beta\alpha\alpha\beta}(q) - \chi^{ch}_{\beta\alpha\alpha\beta}(q)\right),    
\end{align}
is used as a mean of consistency check between single-particle and two-particle objects.
The same relation can also be established in the multi-orbital case~\cite{Zantout2021}. 
Here, we investigate how this consistency changes between the different TPSC extensions.
The comparison is again performed for the model presented in sec.~\ref{sec: suscep}, where we show the results in Table~\ref{tab: internal_check}.

\begin{table}[]
\begin{tabular}{c|c|c|c}
$U/t$     & TPSC & $\langle nn \rangle^{}_{\text{DMFT}}$-TPSC & $\langle nn \rangle^{}_{\text{DMFT}}$-TPSC+$\Sigma_{DMFT}$ \\ \hline \hline
\multicolumn{4}{c}{Orbital 0}                                                                       \\ \hline \hline
$4$ & 6.129e-02& 1.015e-01 & 3.142e-02 \\
$5$ & 9.658e-02& 2.240e-01 & 8.675e-03 \\
$7$ & 2.364e-01& 6.230e-01 & 2.408e-01 \\
\multicolumn{4}{c}{Orbital 1}                                                                       \\ \hline \hline
$4$ & 9.854e-02     & 1.666e-01        &    2.3988e-02         \\
$5$ & 1.467e-01     & 3.460e-01        &    2.9320e-02         \\
$7$ & 3.089e-01     & 7.865e-01        &    2.2499e-01        
\end{tabular}
\caption{Relative error in the $\mathrm{tr}(\Sigma G)$-sum rule obtained from TPSC, $\langle nn \rangle^{}_{\text{DMFT}}$-TPSC, $\langle nn \rangle^{}_{\text{DMFT}}$-TPSC+$\Sigma_{DMFT}$. While the error increases when using only the DMFT double occupations, we observe that this is again compensated when also including the DMFT self-energy, leading to a reduction of the error.}
\label{tab: internal_check}
\end{table}

We observe an increase in relative error between left-hand and right-hand side of the $\mathrm{tr}(\Sigma G)$-sum rule when using $\langle nn \rangle^{}_{\text{DMFT}}$-TPSC instead of TPSC.
This effect can be reversed when the DMFT self-energy is also included.
We attribute this observation to an additional inconsistency when using the DMFT double occupations without inclusion of the DMFT self-energy.
Only when all the correlation effects that are accounted for in DMFT are present on both sides of the sum rule, namely the double occupations and the self-energy, the error decreases again.

\section{Nodal/antinodal spectral function}
\begin{figure*}
    \centering
    \includegraphics[width=\linewidth]{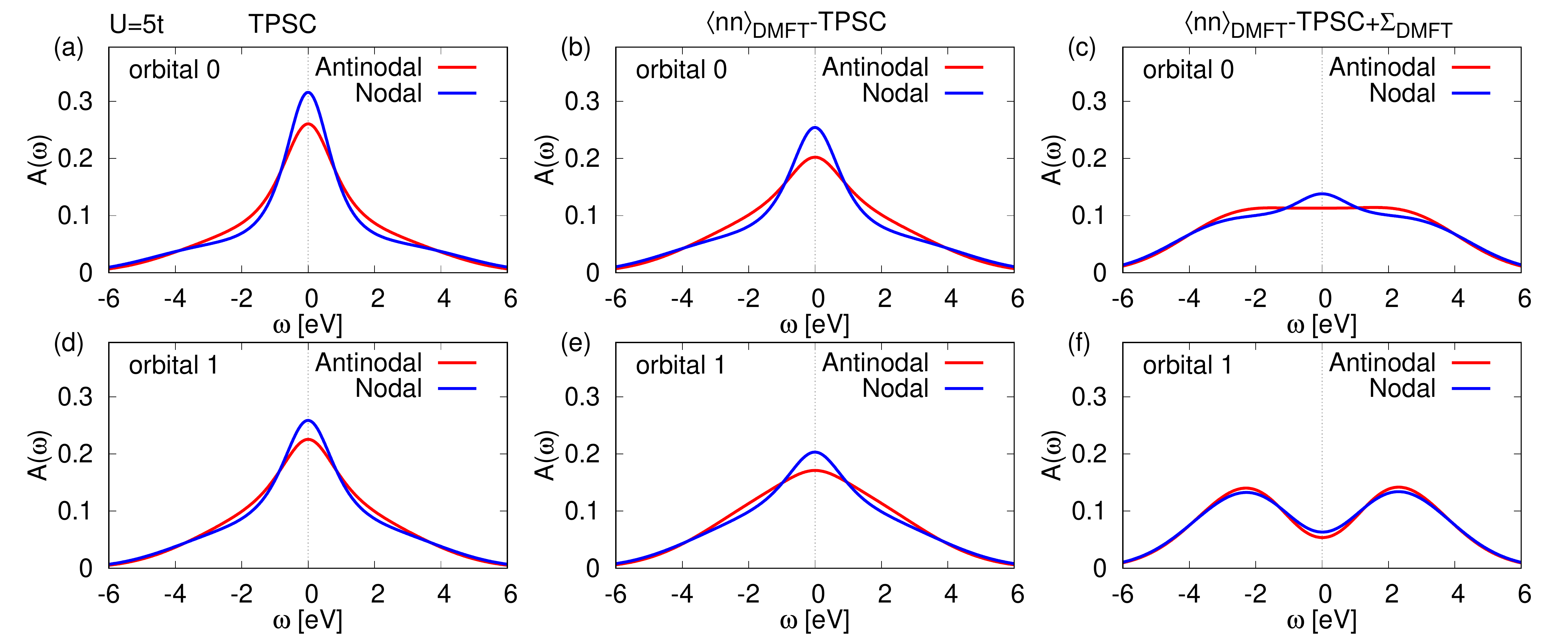}
    \caption{
      Spectral function at the nodal $(\pi/2, \pi/2)$ and antinodal $(\pi,0)$ points, calculated using the three approaches TPSC,  $\langle nn \rangle^{}_{\text{DMFT}}$-TPSC and $\langle nn \rangle^{}_{\text{DMFT}}$-TPSC$+\Sigma_{DMFT}$, for interaction strength $U/t=5$ and temperature $T/t=0.5$. The spectral function at the antinodal point shows a greater suppression of spectral weight than at the nodal point, representative of the tendency to form a pseudogap. Due to a rather high temperature the momentum-dependence is not very pronounced.
    }
    \label{fig:nodal_antinodal_specfunc}
\end{figure*}

To demonstrate the resulting effect of the combined TPSC and DMFT scheme on the Fermi surface and possible emergence of a pseudogap via a the momentum-dependent self-energy we show the analytically continued spectral function at the nodal $(\pi/2, \pi/2)$ and antinodal $(\pi,0)$ points in Fig.~\ref{fig:nodal_antinodal_specfunc} for a fixed value of the interaction $U/t=5$.
We observe a larger suppression of spectral weight at the antinodal point compared to the nodal point, which originates from the larger imaginary part of the momentum-dependent self-energy at the antinodal point, indicative of a tendency to form a pseudogap.
The relative suppression is similar in all schemes, i.e. we find that the momentum-separation of the correlation effects at the nodal and antinodal point is not significantly affected by using the DMFT double occupancies in the DMFT scheme. The replacement of the local TPSC self-energy by the DMFT self-energy does not modify the overall relative momentum-dependence, therefore, in all schemes the momentum-dependence of the self-energy is mostly governed by the original pure TPSC result.

\bibliography{main}

\end{document}